\documentclass[onecolumn,amsmath,amssymb,floatfix,prl]{revtex4}
\usepackage{bm}
\usepackage{amsmath}
\usepackage{epsfig}

\begin{document}
\title{Pearson's random walk in the space of the CMB phases: evidence for parity asymmetry.}
\author{M. Hansen\textsuperscript{1}, A.M. Frejsel\textsuperscript{1}, J. Kim\textsuperscript{1}, P. Naselsky\textsuperscript{1} and F. Nesti\textsuperscript{2}.}

\affiliation{
\textsuperscript{1} Niels Bohr Institute and DISCOVERY center, Blegdamsvej 17, 2100 Copenhagen, 
{\O}, Denmark \\
\textsuperscript{2} University of Ferrara, Physics Department, v. Saragat 1, I-44100 Ferrara, Italy
}

\newcommand{\nc}{\newcommand}
\newcommand{\beq}{\begin{equation}}
\newcommand{\eeq}{\end{equation}}
\newcommand{\be}{\begin{eqnarray}}
\newcommand{\ee}{\end{eqnarray}}
\newcommand{\num}{\nu_\mu}
\newcommand{\nue}{\nu_e}
\newcommand{\nut}{\nu_\tau}
\newcommand{\nus}{\nu_s}
\newcommand{\mnus}{M_s}
\newcommand{\taus}{\tau_{\nu_s}}
\newcommand{\nnt}{n_{\nu_\tau}}
\newcommand{\rnt}{\rho_{\nu_\tau}}
\newcommand{\mnt}{m_{\nu_\tau}}
\newcommand{\tnt}{\tau_{\nu_\tau}}
\newcommand{\rar}{\rightarrow}
\newcommand{\lar}{\leftarrow}
\newcommand{\lrar}{\leftrightarrow}
\newcommand{\dm}{\delta m^2}
\newcommand{\mpl}{m_{Pl}}
\newcommand{\mbh}{M_{BH}}
\newcommand{\nbh}{n_{BH}}
\newcommand{\crit}{{\rm crit}}
\newcommand{\ini}{{\rm in}}
\newcommand{\cmb}{{\rm cmb}}
\newcommand{\rec}{{\rm rec}}

\newcommand{\Odm}{\Omega_{\rm dm}}
\newcommand{\Ob}{\Omega_{\rm b}}
\newcommand{\Om}{\Omega_{\rm m}}
\newcommand{\nb}{n_{\rm b}}
\def\simlt{\lesssim}
\def\simgt{\gtrsim}
\def\Cl{C_{\ell}}
\def\out{{\rm out}}
\def\in{{\rm in}}
\def\mean{{\rm mean}}
\def\zrec{z_{\rm rec}}
\def\zreio{z_{\rm reion}}
\def\wmap{{\it WMAP}} 
\def\planck{{\it Planck}}
\begin{abstract}
The temperature fluctuations of the Cosmic Microwave Background (CMB) are supposed to be distributed randomly in both magnitude and phase, following to the simplest model of inflation. In this paper, we look at the odd and even multipoles of the spherical harmonic decomposition of the CMB, and the different characteristics of these, giving rise to a parity asymmetry. We compare the even and odd multipoles in the CMB power spectrum, and also the even and odd mean angles. \\
We find for the multipoles of the power spectrum, that there is power excess in odd multipoles, compared to even ones, meaning that we have a parity asymmetry. \\
Further, for the phases, we present a random walk for the mean angles, and find a significant separation for even/odd mean angles, especially so for galactic coordinates. This is further tested and confirmed with a directional parity test, comparing the parity asymmetry in galactic and ecliptic coordinates.
\end{abstract}
\maketitle

\section{Introduction}
Gaussianity and statistical homogeneity and isotropy of the CMB are essential for determining the most optimal values of the cosmological parameters in the framework of the $\Lambda CDM$ concordance model. It is remarkable, that recently obtained data by WMAP, BOOMERANG, MAXIMA, CBI etc. clearly demonstrate a consistency between theoretically predicted and observed properties of the CMB anisotropy power spectrum(\cite{WMAP5}, \cite{WMAP3} and \cite{WMAP1}). However, detailed investigations of the statistical properties of the CMB signal reveal a significant deviation from Gaussianity and statistical isotropy of the CMB (\cite{Chiang}, \cite{Coles}, \cite{Vielva}, \cite{Park1}, \cite{Eriksen2}, \cite{Hansen1}, \cite{Larson}, \cite{Copi1}, \cite{Coldspot1}, \cite{Oliveira-Costa1}, \cite{Copi2}, \cite{Land1}, \cite{Schwarz1}, \cite{Naselsky1}, \cite{Eriksen1}, \cite{Hoftuft1}, \cite{Jkim1}), which can be related to the foreground residuals (\cite{Naselsky2}, \cite{Naselsky3}), systematic effects (\cite{Hansen1}), or it could even have primordial origin (for instance the primordial magnetic field, texture or a non-trivial topology of the Universe).

The important task of investigating the departure from statistical isotropy and Gaussianity of the CMB, requires designed tests, which can clarify the possible sources of ``contamination'' in the results, derived from the CMB signal. 
Most of these tests are based on the CMB map, which require a good quality of data reduction, including a high quality of the calibrations, correction of the antenna beam and pixelization uncertainties and a reduction of the residuals of the point-like sources and diffusion foregrounds (and even the incorporation of the uncounted ones, like spinning dust or even the refracting debris of the Kuiper belt \cite{Burigana}). In contrast, recently proposed in \cite{Jkim1}, \cite{Jkim2}, a parity test only requires the power spectrum, which is usually estimated with an accuracy significantly better than the CMB map. Recently, ideas has been put forth, to explain the observed parity asymmetry, see for instance \cite{Zhitnitsky1} and \cite{Zhitnitsky2}. Also \cite{Copi3} has commented on the possibility of biased maps, which might also contribute to parity asymmetry.\\
The idea of our test is based on the analysis of the ratio $g(l)$ of the powers stored in even and odd multipoles. For a statistically homogeneous and isotropic random CMB, $g(l)$ should be equally distributed around $g(l)=1$. 
This assumption is applicable only, if the theoretical power spectrum $C(l)$ of the primordial fluctuations, has a monotonic shape without any oscillations of the corresponding scale $\Delta l$. The monotonic shape of the power spectrum is typical for the simplest models of inflation, when the slow roll approximation is valid, until the epoch of reheating.
However, in more complicated models (for instance in \cite{starobinski}), the slow roll regime of inflation could be broken in the vicinity of some points $\phi\subseteq \{\phi_i\},i=1,2.$, leading to spatial modulations of the power spectrum for adiabatic perturbations of the metric without departure from Gaussianity. In this case parity test can be used as indicator of non-trivial regimes of inflation.
Additionally, we present a directional parity test $G(l)$, similar to to the ratio $g(l)$. This also tests the power stored in even and odd multipoles, but it is based on the directional statistic $D(l, \vec{n})$ originally introduced by \cite{tegmark}. The directional parity test $G(l)$ should also be equally distributed around $1$.

We would like to focus our attention on an extension of the parity test for a very broad range of multipole numbers $l$ for the whole-sky CMB map, taking under 
consideration the phases $\phi_{l,m}$ of the corresponding coefficients $a_{l,m}$ of the spherical harmonic decomposition. 
The phases $\phi_{l,m}$ carries much of the information about the angular distribution of the temperature anisotropy on the sky and possible correlations between different multipoles. 
Moreover, if the parity asymmetry of the power spectrum reflects
possible departure from statistical isotropy of the CMB, the phases $\phi_{l,m}$ should have the same type of asymmetry. Previously, \cite{Raeth1} and \cite{Raeth2} has investigated the phases, and has further found evidence for hemispherical asymmetry \cite{Rossmanith}, \cite{Raeth3}.

The outline of the paper is the following. In Section 1 we introduce the basic characteristics of the power and phases asymmetry. For the phases we will use the 'mean angle' approach, proposed in \cite{Naselsky1} for even and odd multipoles and show that there is strong departure from statistical isotropy. In section 2, we introduce the directional parity test. Section 3 takes a look on the alignment of mean angles, for low $l$. Section 4 is devoted to the random walk of the phases, in both galactic and ecliptic coordinates. In section 5 we present the results of the directional parity test, for ecliptic and galactic coordinates. Finally in section 6 we discuss the findings and draw our conclusions.

\section{Estimators of the parity asymmetry of amplitudes and phases.}
The temperature fluctuations on the sphere $\Delta T (\theta,\phi)$ can be decomposed into spherical harmonics in the following standard way:
\begin{eqnarray}\label{a_lm}
\Delta T (\theta,\phi) = \sum ^{\infty}_{l=0} \sum ^{l}_{m=-l} a_{l,m} Y_{l,m}(\theta,\phi)
\end{eqnarray}
where $\hat{\mathbf n}=(\theta,\phi)$ is a unit vector, pointing in the direction of the polar $\theta$ and azimuthal $\phi$ angles, $ a_{l,m}$ is the coefficient of decomposition: $a_{l,m} = |a_{l,m}| \exp(i \phi_{l,m})$, with $\phi_{l,m}$ as the phase.\\
Under the assumption of total gaussian randomness, as predicted by the simplest inflationary model, the amplitudes $|a_{l,m}|$ follow the Rayleigh's Pdf and phases of $a_{l,m}$ are supposed to be evenly distributed in the range $[0,2 \pi]$ \cite{Bardeen1986}.

\begin{figure}[!htb]
  \begin{center}
    \centerline{\includegraphics[scale=0.5]{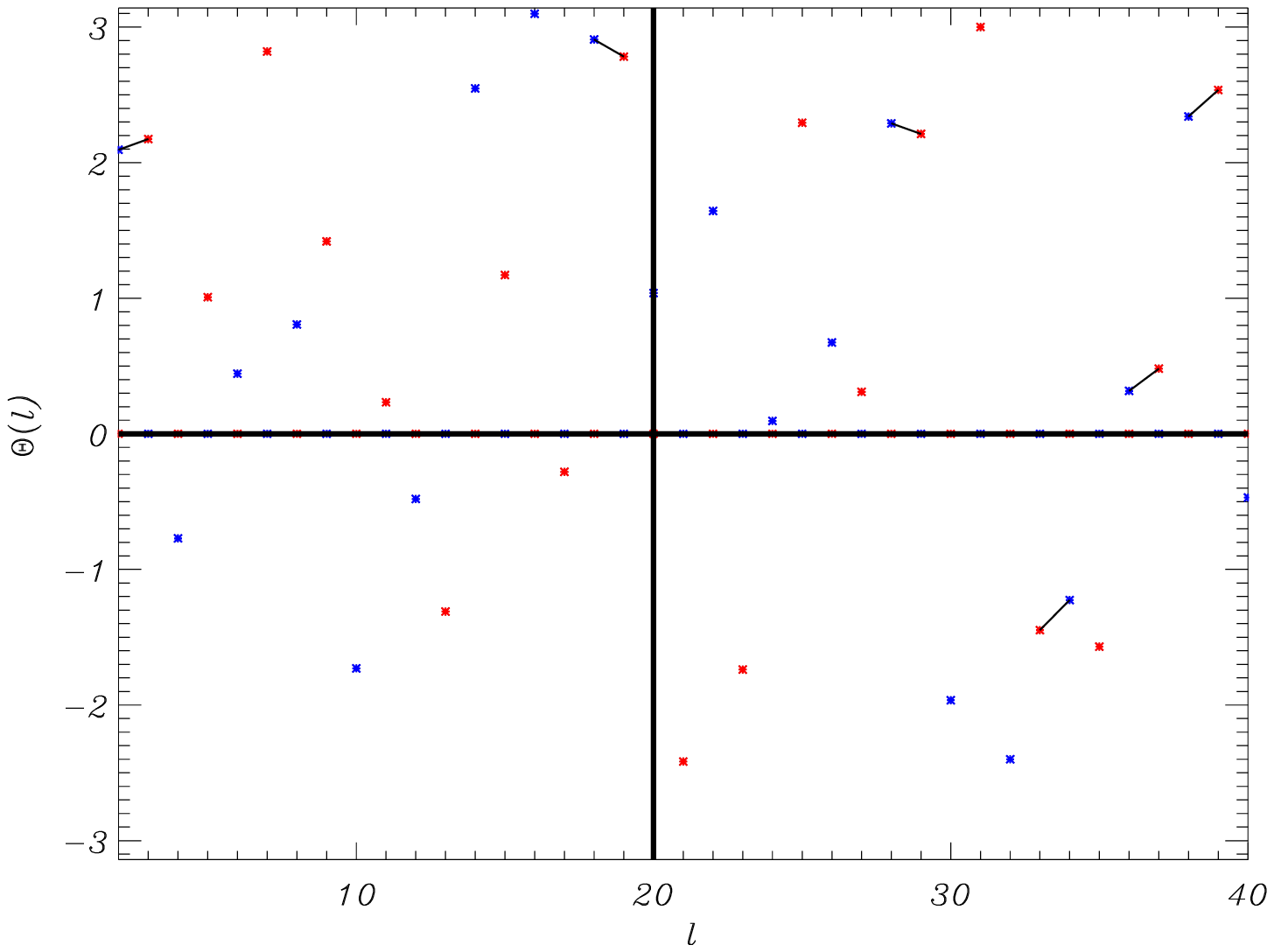}
\includegraphics[scale=0.5]{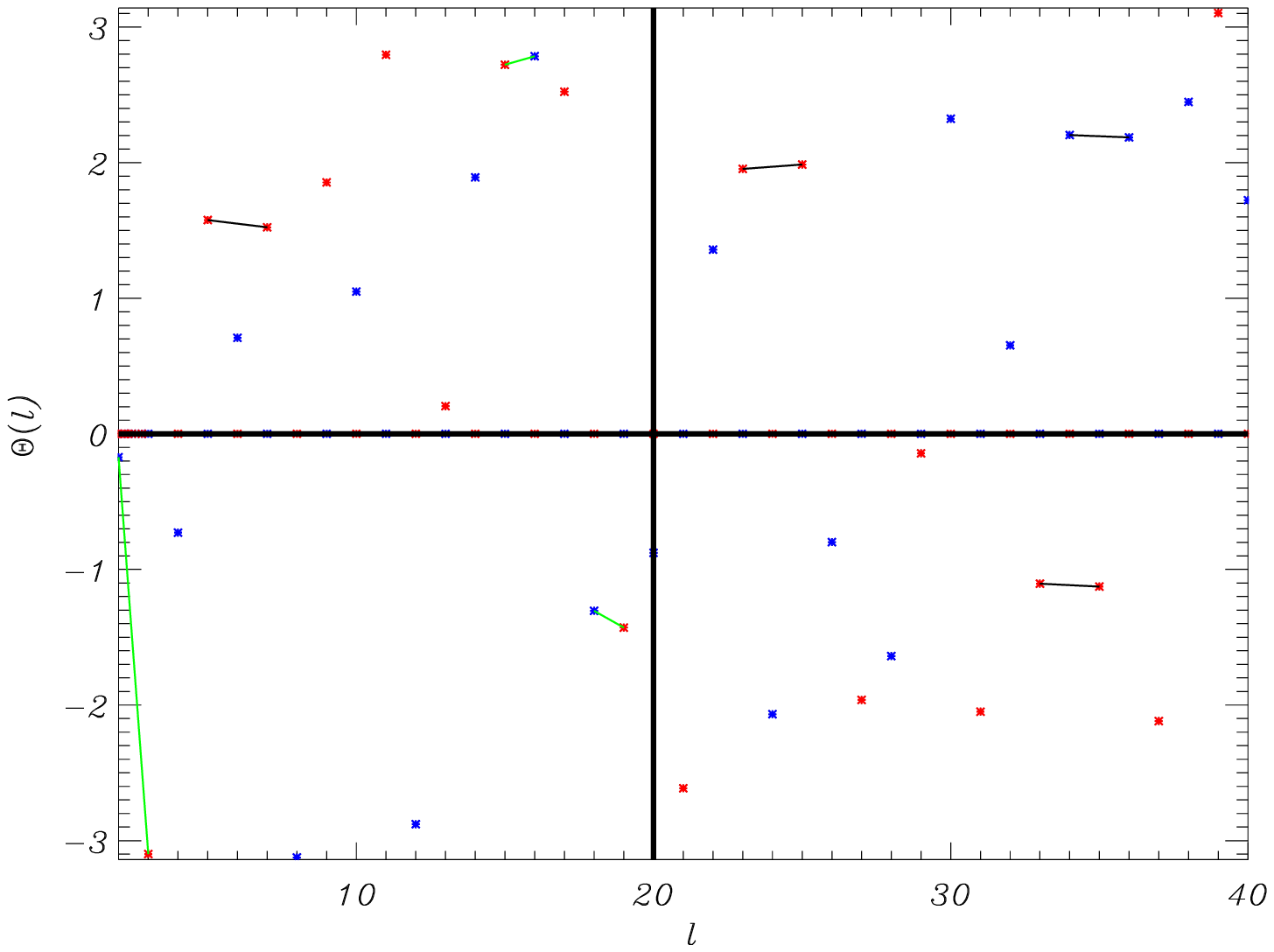}}
    \caption{Left panel. The mean angle for ILC7 phases in galactic coordinates, computed for the first 40 multipoles. Red dots indicates the odd multipoles, and blue dots are for even. A line between two points signify, that the two multipoles have a similar mean angle. Right Panel. The same as left, but for ecliptic coordinates.}
    \label{figure_mean_angle}
  \end{center}
\end{figure}

For any signals $T(\hat{\mathbf n})$ on the sphere, one can define the symmetric ($ T^+(\hat{\mathbf n}) $) and asymmetric ($ T^-(\hat{\mathbf n}) $) components
\begin{eqnarray} 
T(\hat{\mathbf n})=T^+(\hat{\mathbf n})+T^-(\hat{\mathbf n}),
\end{eqnarray}
where
\begin{eqnarray} 
T^+(\hat{\mathbf n})&=&\frac{T(\hat{\mathbf n})+T(-\hat{\mathbf n})}{2}=\sum_l\sum_{m=-l}^{l}a_{l,m}\cos^2(\frac{\pi l}{2})Y_{lm}(\hat{\mathbf n}),\\
T^-(\hat{\mathbf n})&=&\frac{T(\hat{\mathbf n})-T(-\hat{\mathbf n})}{2}=\sum_l\sum_{m=-l}^{l}a_{l,m}\sin^2(\frac{\pi l}{2})
Y_{lm}(\hat{\mathbf n})
\end{eqnarray}
and $Y_{lm}(\hat{\mathbf n})=(-1)^l\,Y_{lm}(-\hat{\mathbf n})$.\\
By definition, $T^+(\hat{\mathbf n})$ and $T^-(\hat{\mathbf n})$ are orthogonal, which means that the average over the whole sphere corresponds to $\langle T^+(\hat{\mathbf n})T^-(\hat{\mathbf n})\rangle=0$ and that the power spectrum of $T(\hat{\mathbf n})$ is given by
\begin{eqnarray} 
 C(l)=\frac{1}{2l+1}\sum_m|a_{lm}|^2=C^+(l)+C^-(l)
\end{eqnarray}
which is associated with power spectrum of even ($C^+(l)=C(l)\cos^2(\frac{\pi l}{2})$) and odd ($C^-(l)=C(l)\sin^2(\frac{\pi l}{2})$) multipoles respectively.
For the concordance $\Lambda$CDM cosmological model with initial Gaussian adiabatic perturbations, we do not expect any features distinct between even and odd multipoles. However, taking the WMAP7 data as a platform for investigation of the parity of the CMB, there have been reported power contrast between even and odd multipoles of WMAP $TT$ power spectrum (\cite{Universe_odd}, \cite{Jkim1}, \cite{odd_origin}, \cite{odd_bolpol}, \cite{WMAP7:anomaly}).
The corresponding estimator for ``even and odd'' asymmetry of the CMB power spectrum is given by\cite{Jkim1}
\begin{eqnarray}
 g(L)=\frac{\sum_{l=2}^L l(l+1)C^+(l)}{\sum_{l=2}^L l(l+1)C^-(l)}
\label{par_est}
\end{eqnarray}
The $g(l)$-statistic is a cumulative statistic, but it is the most practical approach to testing for asymmetry between even and odd multipoles. The problem is, that we cannot make a comparison for a single l, as either the odd or even component would be zero. An alternative could be to compare multipoles of a separation of $1$ for instance, i.e. compare an odd multipole with its even neighbor, however we would have to decide what neighbor to compare to ($l$ and $l+1$, or $l$ and $l-1$), and the choice would be an arbitrary one. In \cite{Universe_odd} they performed a test for neighboring $C_{l}$'s, comparing $l$ and $l-1$. We tested that method also, and found that it did not work as well as eq. \ref{par_est} for our specific purposes, and thus we kept the $g(l)$-statistics as introduced by \cite{Jkim1}.\\a
At lowest multipoles ($2\le l\le 22$), there is reported odd multipole preference (i.e. power excess in odd multipoles and deficit in even multipoles) (\cite{Universe_odd}, \cite{Jkim1}, \cite{odd_origin}, \cite{odd_bolpol}).
Note that due to rotational invariance of $g(l)$, the corresponding parity asymmetry does not depend on the reference system of coordinates. 
However, additional information about the statistical properties of the temperature fluctuations is associated with the odd and even components of the phases $\phi_{l,m}$, which are rotationally non-invariant. This is why the analysis of the CMB phases $\phi_{l,m}$, for different systems of coordinates seems to be extremely important for determining the origin of the CMB parity asymmetry.\\
The phases carry information about the orientation of the spherical harmonic components, to make them fit the observed microwave radiation. In order to check the statistical anisotropy of the phases $\phi_{l,m}$ for $l=odd$ and $l=even$, we will use the trigonometric moments:
\begin{eqnarray}\label{si_org}
\textbf{Si}(l) = \frac{1}{2l +1} \sum^{l}_{m=-l} \sin(\phi_{l,m}),\hspace{0.5cm}
\textbf{Ci}(l) = \frac{1}{2l+1} \sum^{l}_{m=-l} \cos(\phi_{l,m})
\end{eqnarray}
The mean angle of the given multipole, $l$, is the arc+tangent to the ratio between the average $sin$ and $cos$ value for the multipole:
\begin{eqnarray}\label{mean_angle}
\Theta(l) = \arctan\left(\frac{\textbf{Si}(l)}{\textbf{Ci}(l)}\right)
\end{eqnarray}
The reason for the above procedures for finding the mean angle, is that we are dealing with circular data and thus a simple average over the phases is not enough. Instead we must convert the angles to points on the unit circle via $sin$/$cos$, and find the arithmetic mean of these points (Eq. \ref{si_org}). This mean is then converted back to polar coordinates via arctan (Eq. \ref{mean_angle}) to give us the mean angle for the considered multipole. Thus we have greatly compacted the information of the CMB phases, by creating a single mean angle, for each multipole. Further details about the procedures above can be found in \cite{fisher}.\\
We would expect, that the mean angles for the multipoles would distribute themselves evenly in the range from $-\pi$ to $\pi$, as required by a Gaussian random field \cite{ferreira}. Moreover, for a statistical homogeneous and isotropic random Gaussian field generated by primordial perturbations, the uniformity of the distribution of phases leads to a uniformity of the distribution of mean angles $\Theta^{\pm}(l)$ for even and odd $l$.\\
For testing the equivalence of the odd and even multipoles in the space of phases, we will use the following estimators: 
\begin{eqnarray}\label{m_a}
R^{\pm}(l) = \frac{1}{l-2}\left\{[r^{\pm}_s(l)]^2+[r^{\pm}_c(l)]^2\right\},\hspace{0.5cm} r^{\pm}_s(l_{max})=\sum_{l=2}^{l_{max}}\sin\Theta^{\pm}(l),\hspace{0.5cm}
r^{\pm}_c(l_{max})=\sum_{l=2}^{l_{max}}\cos\Theta^{\pm}(l),\hspace{0.5cm}
l\ge 3.
\end{eqnarray}
where $\Theta^+$ is the mean angle for even multipoles, $\Theta^-$ is the mean angle for the odd, and the corresponding sum is defined for $n=2k$ and $n=2k+1,k=1,2...$ for $\Theta^+$ and $\Theta^-$. After simple algebra, Eq(\ref{m_a}) gives us the following expression for the function $R^{\pm}$:
\begin{eqnarray}\label{R_a}
R^{\pm}(l) =\frac{1}{l-2}\sum_{n,n'}^l\cos\left(\Theta^{\pm}(n)-\Theta^{\pm}(n')\right)=\nonumber\\
=\frac{1}{2}\left(\delta_{l,2k}+\frac{l-1}{l-2}\delta_{l,2k+1}\right) +\frac{1}{l-2}\sum_{n,n'\neq n}^l\cos\left(\Theta^{\pm}(n)-\Theta^{\pm}(n')\right);
\end{eqnarray}
From Eq(\ref{R_a}) one can see, that for non-correlated mean angles $\Theta^{\pm}(n),\Theta^{\pm}(n')$ the last term can be estimated
as $l^{-2}$ for $l\gg 2$, and thus asymptotically $R^{\pm}(l)\rightarrow \frac{1}{2}$ (see for details \cite{Naselsky4}).
Actually, $R^{\pm}$ is well known in the theory of statistical analysis of circular data (see for instance, \cite{Mardia}). The important generalization of this test, is that it will be applied for odd and even multipoles separately, describing a random walk of the mean angles in the space of phases.\\
If the mean angles for given multipoles $\Theta^{\pm}(n),\Theta^{\pm}(n')$ are close to each other ($\cos\left(\Theta^{\pm}(n)-\Theta^{\pm}(n')\right)\sim \pm 1$), the second term in Eq(\ref{R_a}) is comparable with the first one. For the WMAP CMB map this effect is well known for the quadrupole and the octopole, as alignment. However, in the next section, we will show that the quadropole/octopole alignment is not unique. It is detected for $\Delta l=1$ mean angles (even-odd multipole correlations) in the galactic system of coordinates, and especially, for $\Delta l=2$ in ecliptic coordinates (see fig \ref{figure_mean_angle}).
Thus the parity test of the mean angles $\Theta^{\pm}(l)$ needs to be taken under consideration as a complimentary test to the parity asymmetry of the power spectrum, in order to clarify the origin of these anomalies in the CMB. 

\section{The comparison of the directional parity tests.}
Historically, investigation of the low multipole anomalies of the CMB signal by rotationally non-invariant (directional) statistic was proposed in \cite{tegmark}. The estimator of the preferable direction $\vec{n}$, which maximize the angular momentum($\textbf{L}$) dispersion, can be defined as 
\begin{eqnarray}
 D(l,\vec{n})=\langle\Delta T(\vec{n})((\vec{n}\textbf{L})^2\Delta T(\vec{n})\rangle=\sum_m m^2|a_{l,m}(\vec{n})|^2
\label{teg}
\end{eqnarray}
where the $\vec{n}$ is selected, so that the term $(\vec{n}\textbf{L})$ has the maximum value.\\
Actually, one can define the same estimator, as Eq(\ref{teg}), for the map of gradient in azimuthal direction $\Delta T_{\phi}=\frac{\partial}{\partial\phi}\Delta T(\vec{n})$ for a given system of coordinates with $z$-axis along the vector $\vec{n}$.
Then, based on the Eq(\ref{teg}), we can find the direction of maximal parity asymmetry of the CMB, for the considered directions, specified by $\vec{n}$, just by replacing $l(l+1)C(l)$ with this new
$D(l,\vec{n})$ in Eq(\ref{par_est}), to get the following:
\begin{eqnarray}
 G(L)=\frac{\sum_{l=2} ^L D^{+}(l,\vec{n})}{\sum_{l=2} ^L D^{-}(l,\vec{n})}
\label{G(l)}
\end{eqnarray}
In other words, an estimator precisely as Eq(\ref{par_est}), just using $D(l,\vec{n})$ instead of $C(l)$, for selected values of $\vec{n}$. In this work, we only test for two $\vec{n}$'s, the one corresponding to galactic coordinates, and the one corresponding to ecliptic.

For implementation of the parity test we use the Internal Linear Combination (ILC7) map. This is, according to the WMAP-team, the most precise data, when one is interested in multipoles $l \leq 32$, and it is indeed the very same data, that the WMAP-team uses for finding cosmological parameters \cite{WMAP7_derived_parameters}. 
\begin{figure}[!htb]
  \begin{center}
    \centerline{\includegraphics[scale=0.6]{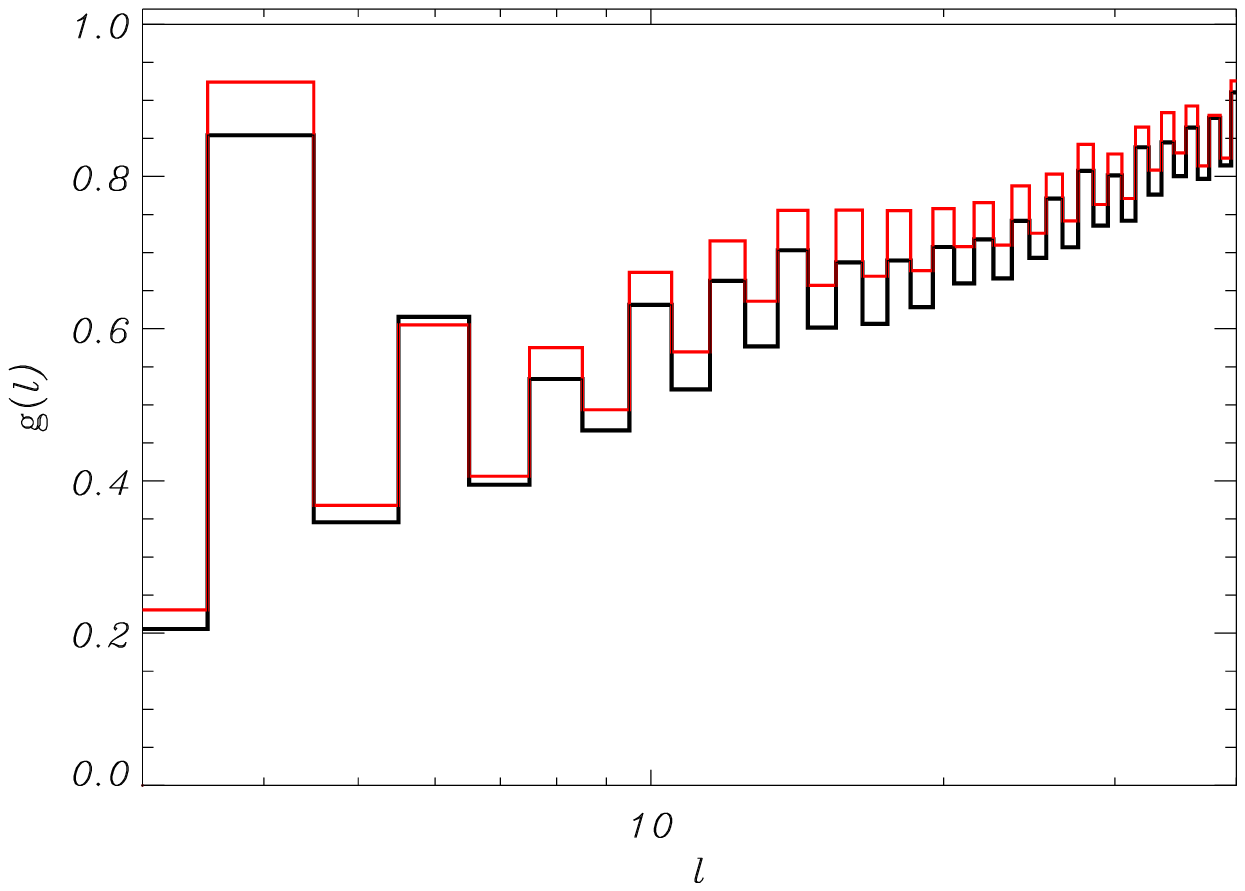}
\includegraphics[scale=0.6]{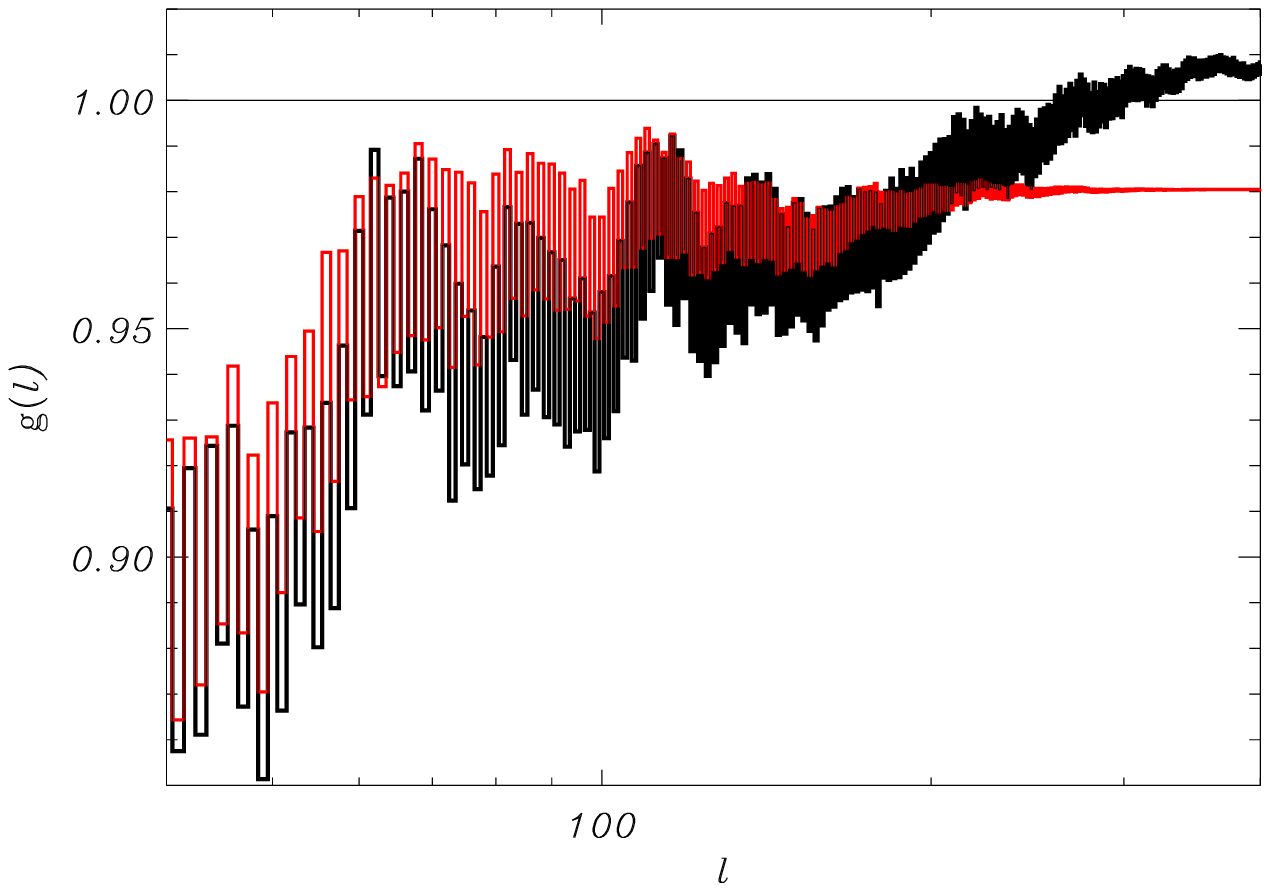}}
    \caption{Left panel. The parity parameters $g(l)$, computed for the range of multipoles $l\le 40$. The thick black line corresponds to WMAP7 power spectrum from the incomplete sky, the red line corresponds to the ILC7. Right panel. The same as left, but for $40\le l \le 400$. The red line represent the ILC7.}
    \label{figure_par}
  \end{center}
\end{figure}

\begin{table}[htb!]
\centering
\caption{Estimator for ILC 7 aligned multipoles at $2\le l\le 40$}
\begin{tabular}{cccccc}
\hline 
    $\cos(\Theta(l) - \Theta(l+1))$ & & $\cos(\Theta(l) - \Theta(l+2))$& &$\cos(\Theta(l) - \Theta(l+1))$ \\
 galactic coordinates & &  ecliptic coordinates  & &ecliptic coordinates\\
\hline
$l=2,3$  & $0.9714$ &$ l=5,7$ & $0.9986$& $l=15,16$& $0.9980$ \\ 
$l=18,19$  & $0.9947$ &$l=23,25$ & $0.9995$& $l=18,19$& $0.9924$ \\ 
$l=28,29$  & $0.9978$ &$l=33,35$ & $0.9998$ \\ 
$l=33,34$ & $0.9477$ &$l=34,36$ & $0.9999$ \\ 
$l=36,37$  & $0.9299$ & & \\ 
$l=38,39$  & $0.9485$ & &\\
\hline
\end{tabular}
\label{EstimatorVal1}
\end{table}

\begin{figure}[!htb]
  \begin{center}
    \centerline{\includegraphics[scale=0.5]{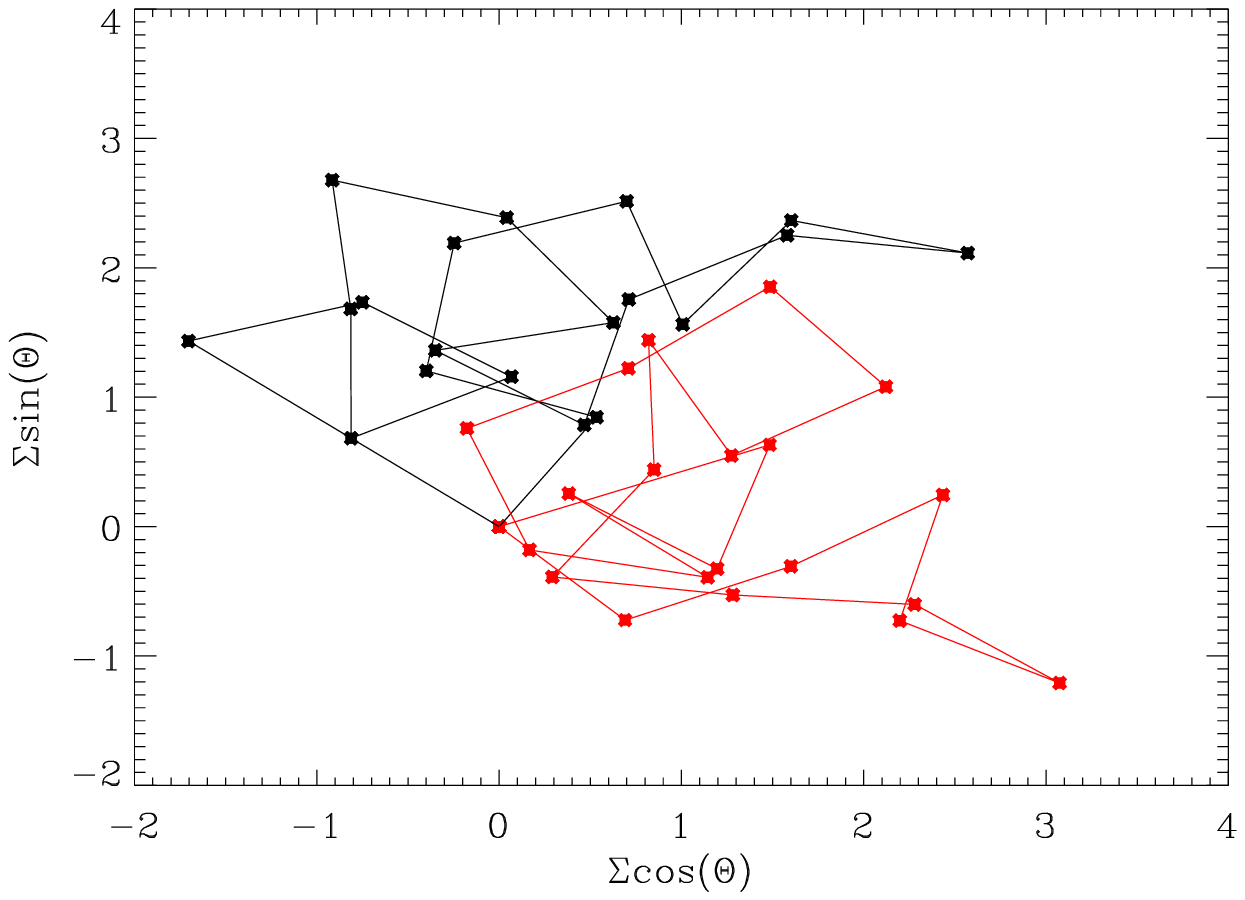}
\includegraphics[scale=0.5]{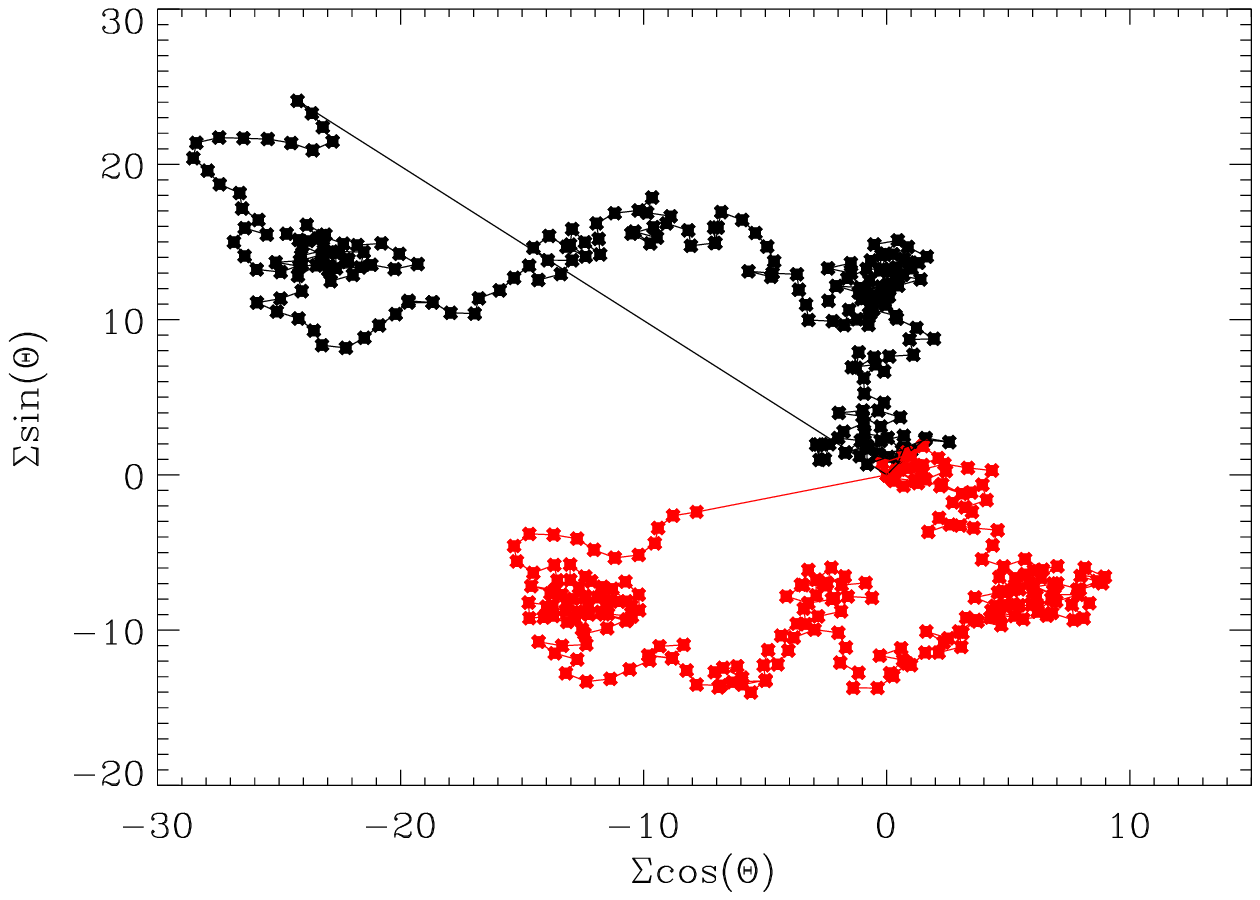}}
    \centerline{\includegraphics[scale=0.5]{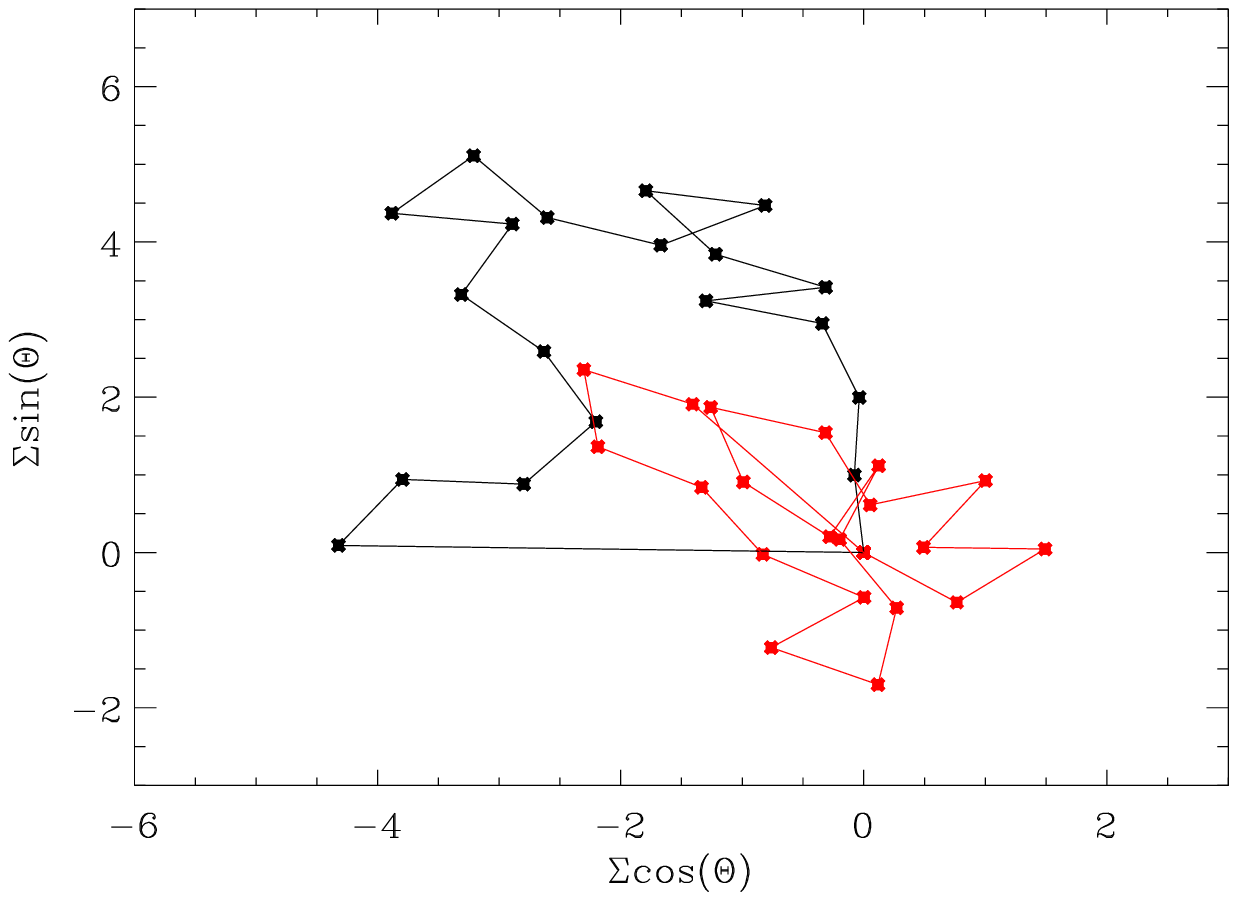}
\includegraphics[scale=0.5]{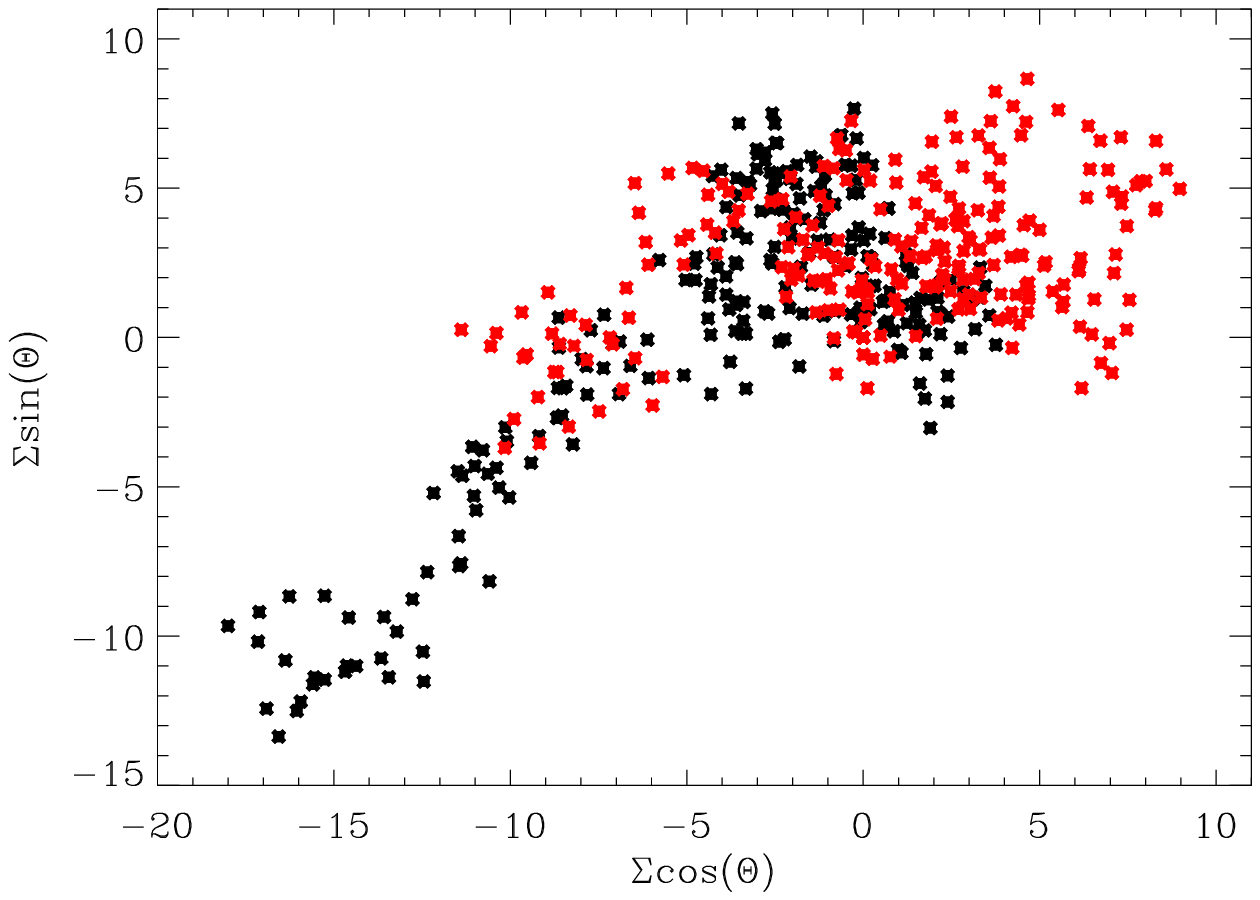}}
    \caption{Top panel. The parameter $r^-(l)$ (the black dots) for odd multipoles and $r^+(l)$ (the red dots) for even ones in Galactic coordinates. Left graph is for the first 40 multipoles, right is for $l$ up till 500. Bottom panels. The same as the top panels, but for ecliptic coordinates.}
    \label{figure_mean_walk}
  \end{center}
\end{figure}

\begin{figure}[!htb]
  \begin{center}
    \centerline{\includegraphics[scale=0.5]{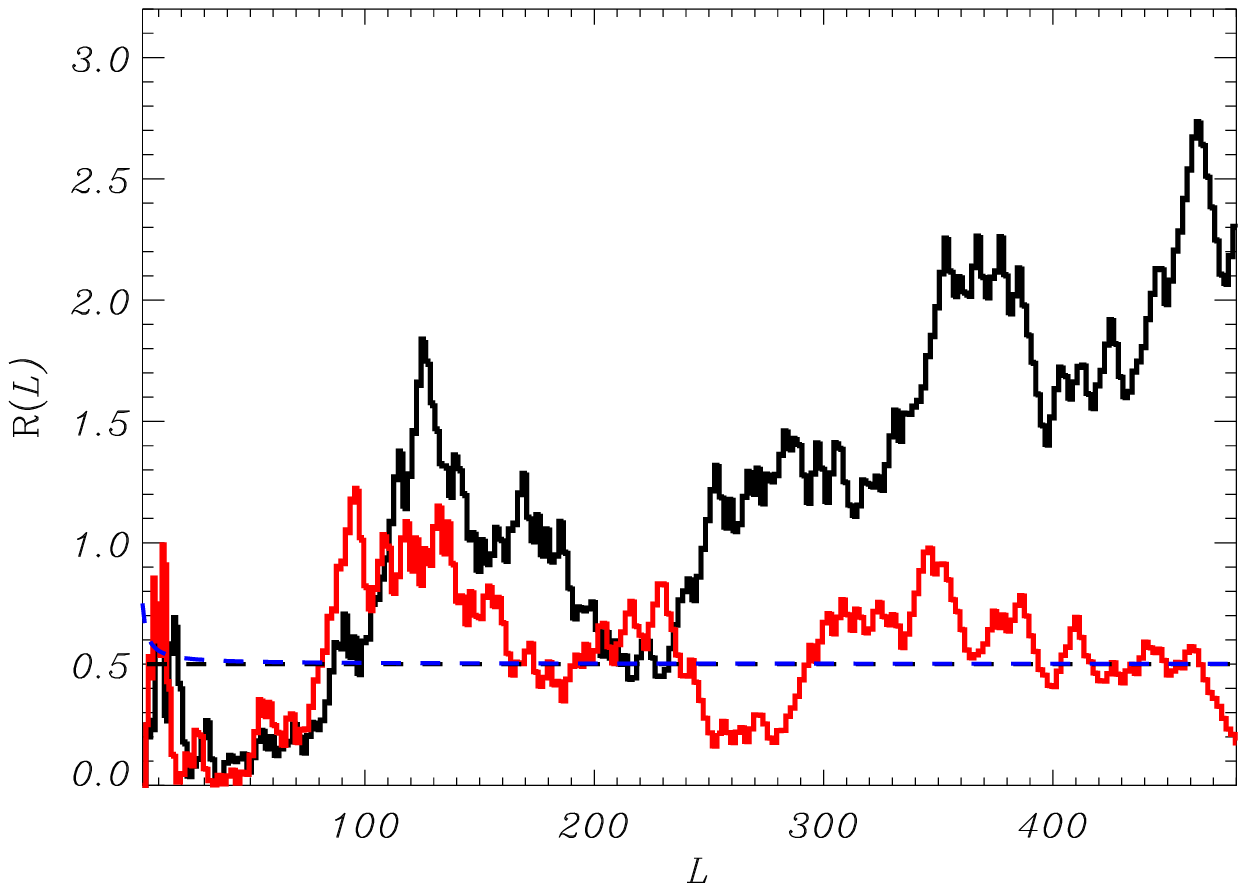}
\includegraphics[scale=0.5]{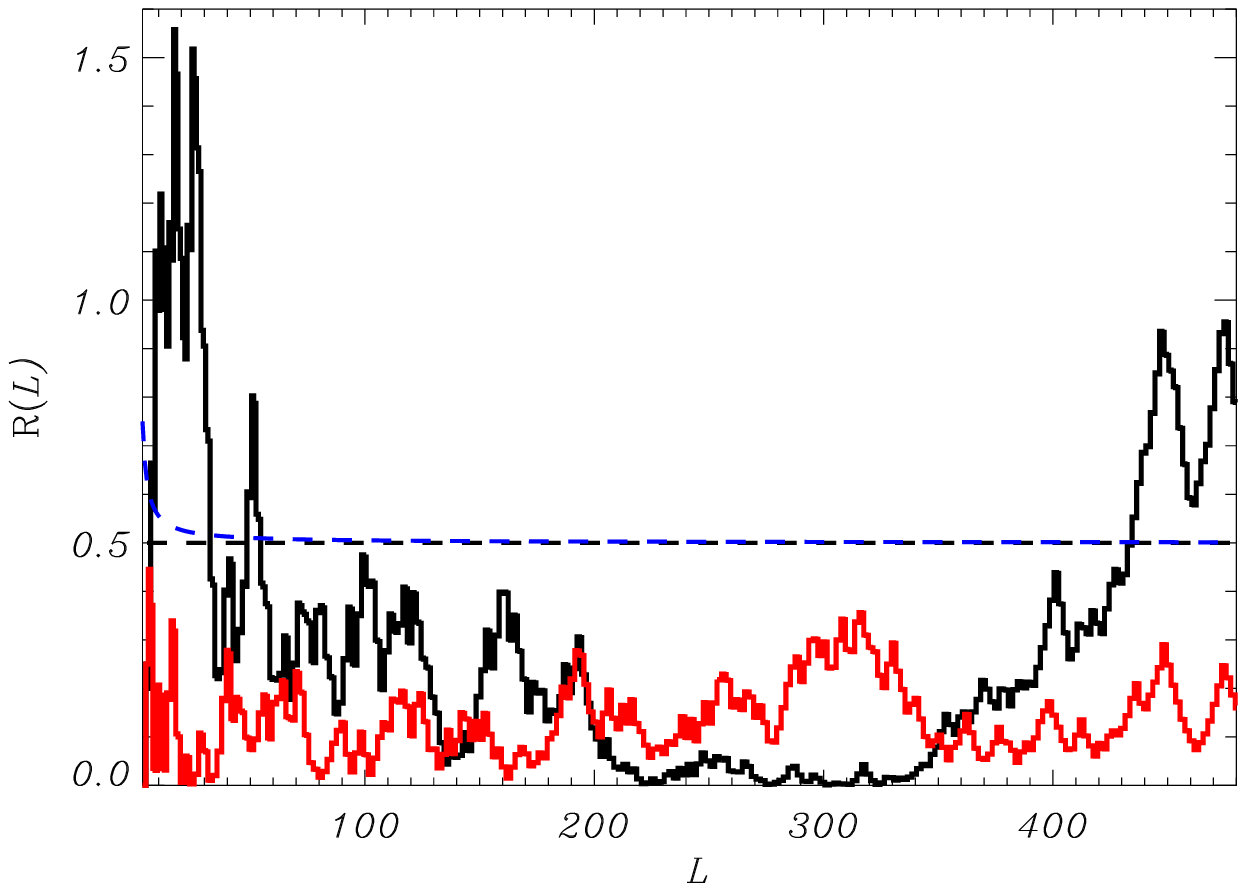}}
    \centerline{\includegraphics[scale=0.5]{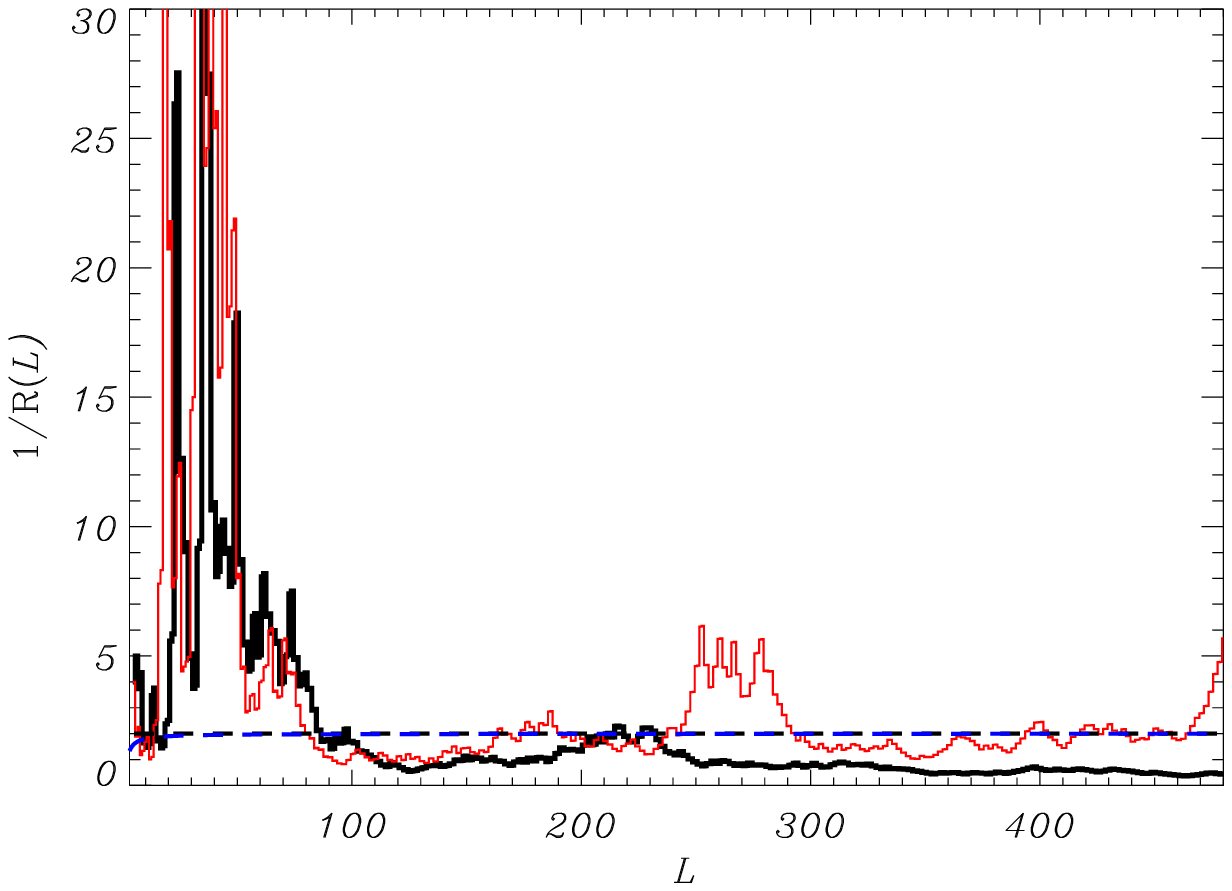}
\includegraphics[scale=0.5]{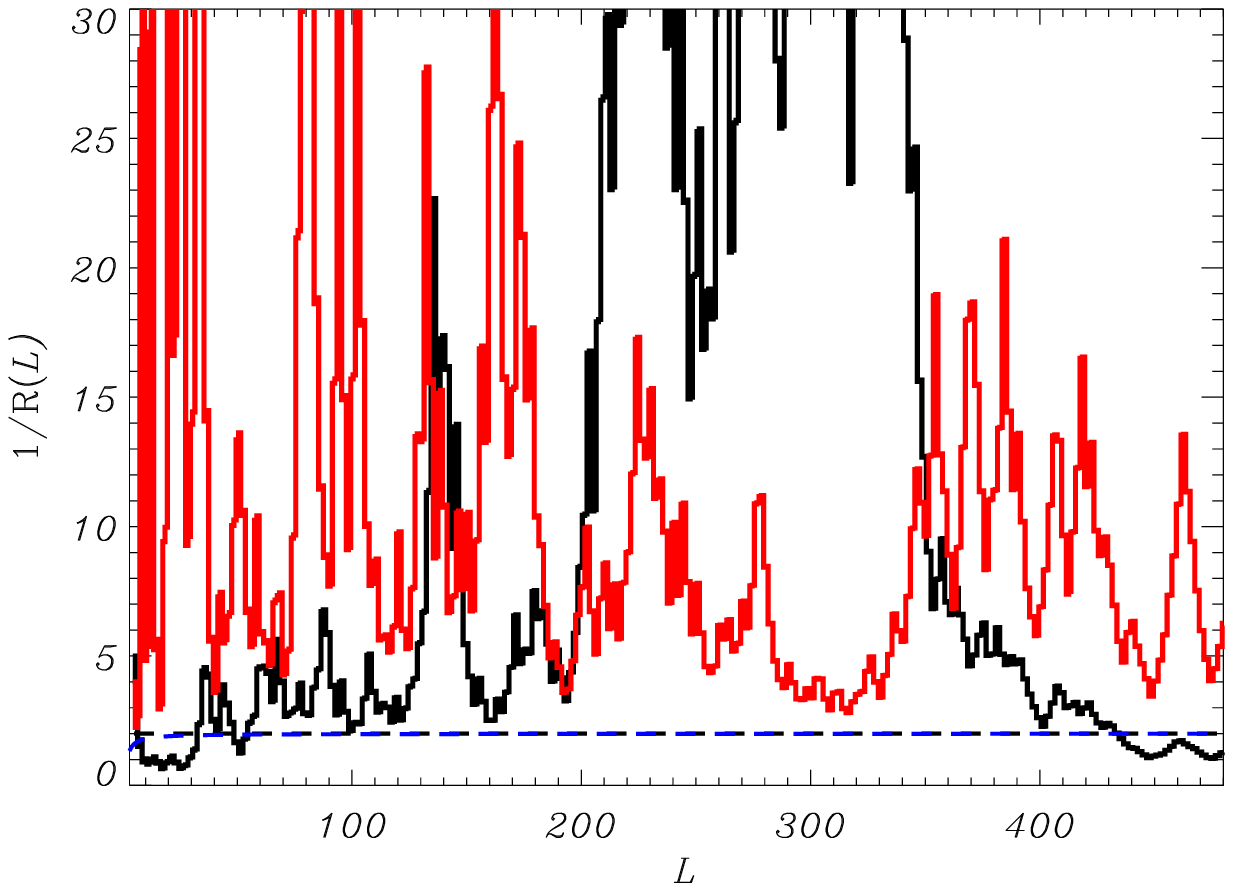}}
    \centerline{\includegraphics[scale=0.5]{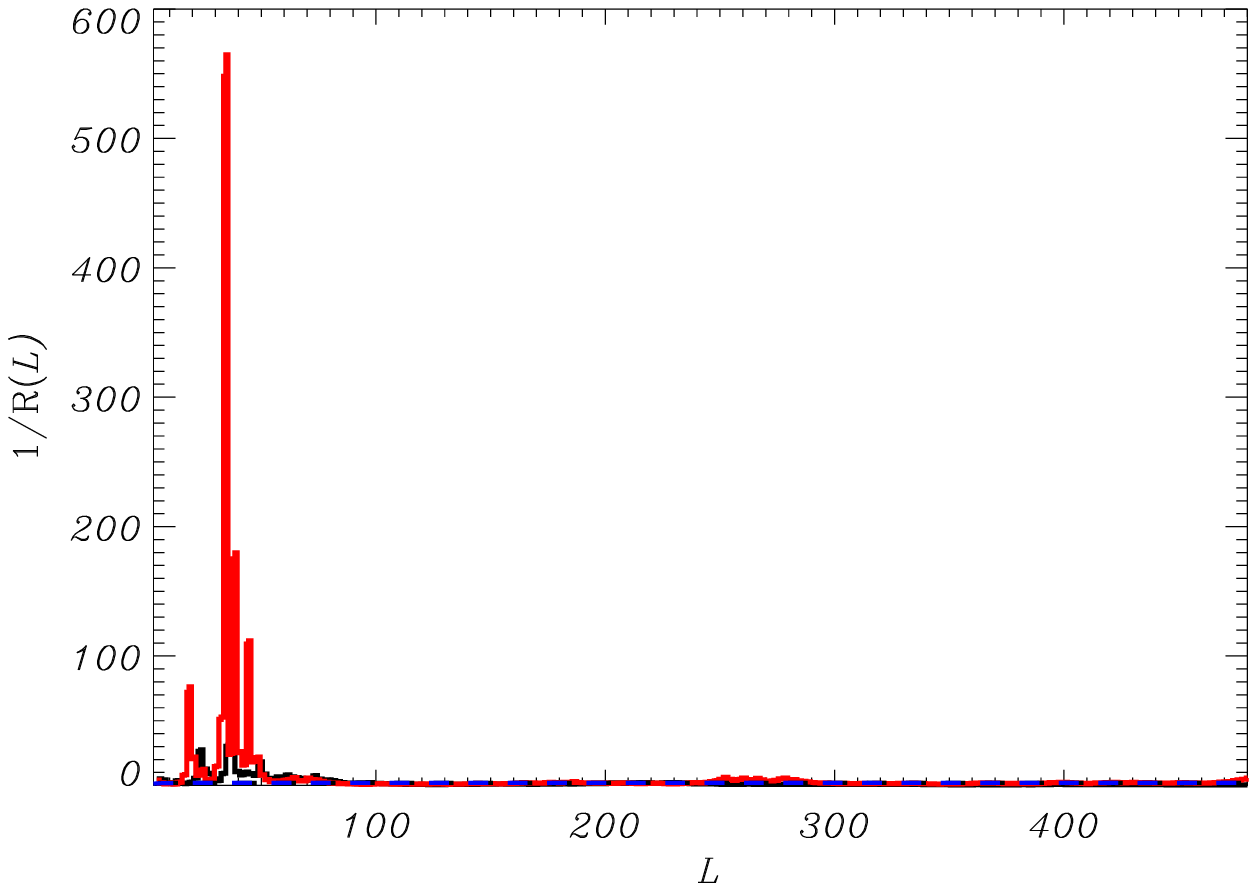}
\includegraphics[scale=0.5]{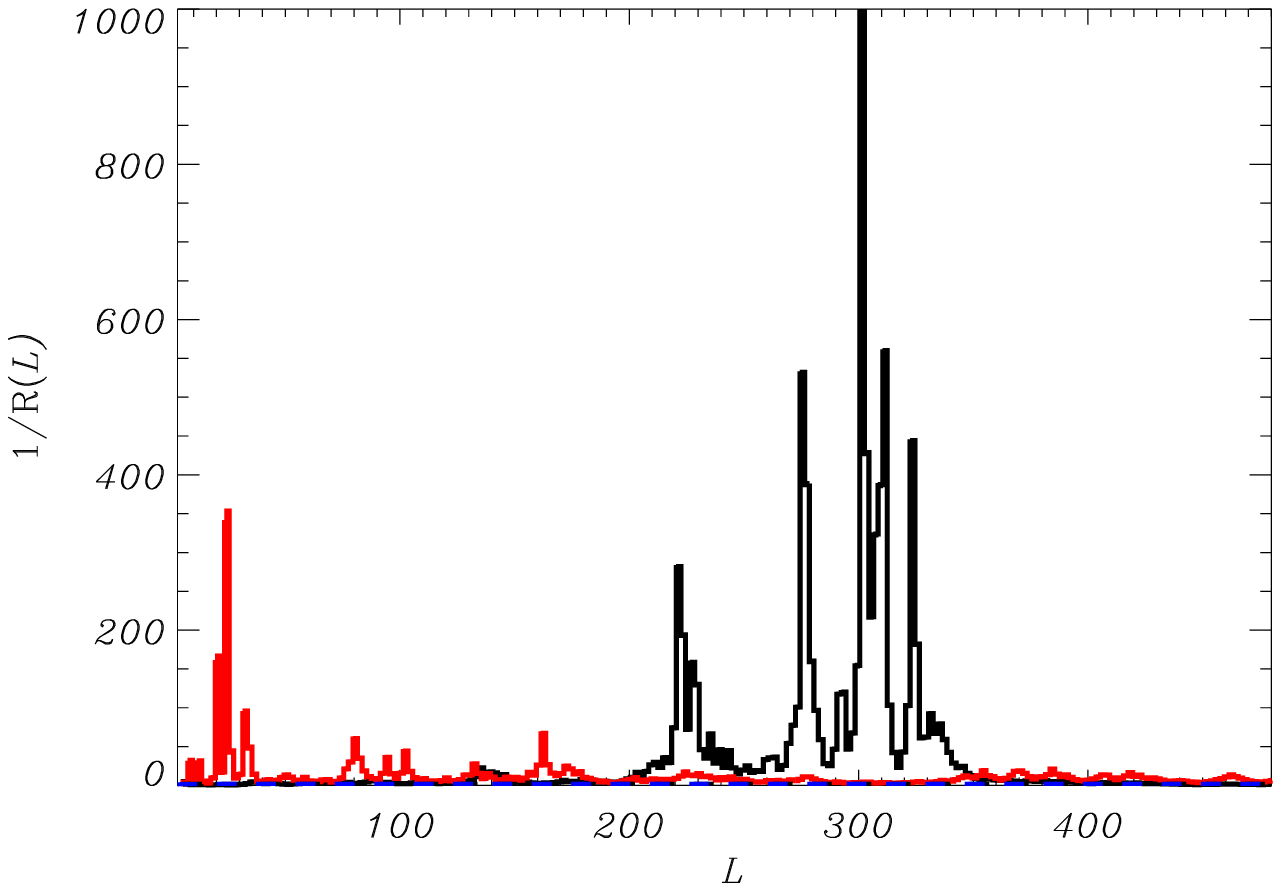}}
    \caption{Left column.The parameters $R^-(l)$ (top panel, the black line) for odd multipoles and $R^+(l)$ (top panel, the red line) for even ones versus $l$ in galactic coordinates. Right column. The same as left, but for ecliptic coordinates. The black dash line is the asymptotic $R^+(l=even)=0.5$, the blue dash line is for $R^-(l=odd)=\frac{l-1}{2(l-2)}$. Second and third from the top are for $1/R^-(l)$ and $1/R^+(l)$, at varying scales.}
    \label{figure_mean_w}
  \end{center}
\end{figure}

\section{Statistic of the parity parameter and the mean angles.}
For illustration of the method we select the WMAP 7 year data from $ http://lambda.gsfc.nasa.gov/$, and following to the procedure above, we calculate the
parity parameter $g(l)$ for $l\le 400$, and the mean angles for galactic, and then for ecliptic coordinates, bearing in mind that the most statistically significant result for $g(l)$ take place for $2\le l\le 22$. The results for the power spectrum are plotted in Fig \ref{figure_par} in comparison with parity parameter $g(l)$ taken from the
incomplete sky (the WMAP power spectrum).
 Since $g(l)< 1$ for all models at this range of multipoles, the power of the CMB $C(2\le l \le 40)$ has deficit of $C(l=even)$ modes in comparison with $C(l=odd)$ and vice versa. To clarify
the origin of the parity asymmetry, we need to look closely at the complimentary test: the parity of mean angles.

In the theory of circular statistics of phases it is well known, that for uniform and non-correlated $\phi_{l,m}$, the mean angles $\Theta^{\pm}(l)$ should be uniform and non-correlated as well. Similar to Eq(\ref{mean_angle}) we can define the mean angles $\xi^{\pm}(l\subseteq \Lambda)$ for a given range of multipoles, within some interval $\Lambda$.
Namely
\begin{eqnarray}
\xi^{\pm}(l\subseteq \Lambda)\equiv \tan^{-1}\left(\frac{\sum_{l\subseteq \Lambda}\sin\Theta^{\pm}(l)}{\sum_{l\subseteq \Lambda}\cos\Theta^{\pm}(l)}\right)
\label{rot9}
\end{eqnarray}

The properties of the mean angle at any system of coordinates, depend on the correlation between the phases for a given multipole $l$ and the different $m$ for this multipole. In a separate paper, we will perform a general analysis of the most preferable direction of the phase correlations trough the mean angle statistics, trying to systematize the most peculiar directions. Here, for simplicity we will test the mean angle of the phases for even and odd $\Theta^{\pm}$ in galactic and ecliptic coordinates.

For the ILC 7 map in the Galactic coordinates at $2\le l\le 40 $ the odd multipoles (red dots in the figure \ref{figure_mean_angle}), have a mean angle of $\xi^-=1.2238$ rad in the range $\Lambda=2-20$. The distribution of the even multipoles is characterized by the mean angle $\xi^+=1.1919$ rad. Note that without separation of the phases into odd and even one, we have for the mean angle $\xi =1.2080$ rad if $2\le l\le 20$. 
For the range $20\le l \le 40$ we have $\xi^-=2.94286$, $\xi^+=0.6783$ and $\xi=1.78893$ rad.
Thus, the cross-correlations between odd and even modes for this range $\Lambda=2-20$ are given 
by $K=\cos(\xi^+-\xi^-)=0.99949$ and $K=-0.639$ for $\Lambda=20-40$. 
For the ecliptic system of coordinates we have $\xi^-=1.06046$, $\xi^+=2.0799$, $\xi=1.6152$ rad. and $K= 0.5238$, where $2\le l \le 20$, and $\xi^-=-0.08576$, $\xi^+=-3.1331$, $\xi=-0.32940$ rad. and $K= -0.99556 $ for $20\le l\le 40$. 
Here we would like to direct attention to the remarkable similarity of the directions
$\xi^-$ and $\xi^+$ for the ILC7 in Galactic coordinates, at the range $2\le l \le 20$. Based on uniformity of the distribution of the mean angles, the difference $\xi^--\xi^+$ should be uniformly distributed within the interval $-\pi,\pi$. The probability that both these angles differ by an angle $\Delta=|\xi^--\xi^+|$ is given by $P=\Delta/(2\pi)\simeq 0.0051$. Note that this probability is close to the probability of getting the parity asymmetry of the power spectrum $0.0044$ for the same range of multipoles, presented in 
\cite{odd_origin}. Both of these directions are highly aligned with the mean angle of the first 19 multipoles (without separation into odd and even multipoles) in the Galactic system of coordinates.

The alignment between $\xi^-,\xi^+$ for $2\le l \le 20$ is not a unique feature for the mean angles of the ILC 7 phases. As it is seen from Fig\ref{figure_mean_angle}, there exists a coupling between even and odd mean angles, most obviously known for $l=2$ and $l=3$\cite{Copi1}. In Table \ref{EstimatorVal1} we show the most interesting alignments between even and odd, odd-odd and even-even multipoles, for which the corresponding correlations exceeds the factor $90\%$.

\section{Pearson's random walk in the domain of mean angles.}
In Fig\ref{figure_mean_walk} we illustrate Pearson's random walk in Galactic and ecliptic coordinates, in the space of mean angles, in term of the parameters
$r^{\pm}_s$ versus $r^{\pm}_c$, for the ILC 7 map at the range of multipoles $2\le l\le 40$.
Note, that the random walk is of course dependent on the chosen coordinate system, so we expect different results from ecliptic and galactic coordinates.
The upper limit $l=40$ has been chosen, in order to avoid contribution from instrumental noise into the phases of the CMB signal. Also included, is a plot with $l$ up till $500$.
The figure shows the separation of the even and odd mean angles in the space of phases. For ecliptic coordinates this tendency
is quite weak, and the corresponding probability to get this effect from a random field is in the order of $1.5\sigma$, where
$\sigma=\sqrt{l/2}\simeq 6.324$. 
 
In galactic coordinates the statistical significance of the separation of even-odd modes exceeds $3\sigma$ threshold.
More qualitatively the contribution from different multipoles into the even-odd random walk is illustrated in Fig.\ref{figure_mean_w}.
From this figure one can see, that a major contribution to the $R^-(l)$-parameter is given by the mean angles at $5\le l\le 30$, where
$R^-(l)$ exceeds the $2\sigma$ threshold. The even multipoles, in contrast with the odd ones, are characterized by very small values of the
$R^+(l)$-parameter, which indicate the presence of cross-correlations between even-even modes.\\
It should be noted, that we expected values of $R^-(l)$ and $R^+(l)$ around $0.5$ ( and thus $1/R\simeq 2$ for the $R^{-1}$-graph). That means that such very small values of $R$ are as statistically unlikely, as very high ones.

\section{Directional $G(l)$-test.}

\begin{figure}[!bth]
  \begin{center}
    \centerline{\includegraphics[scale=0.6]{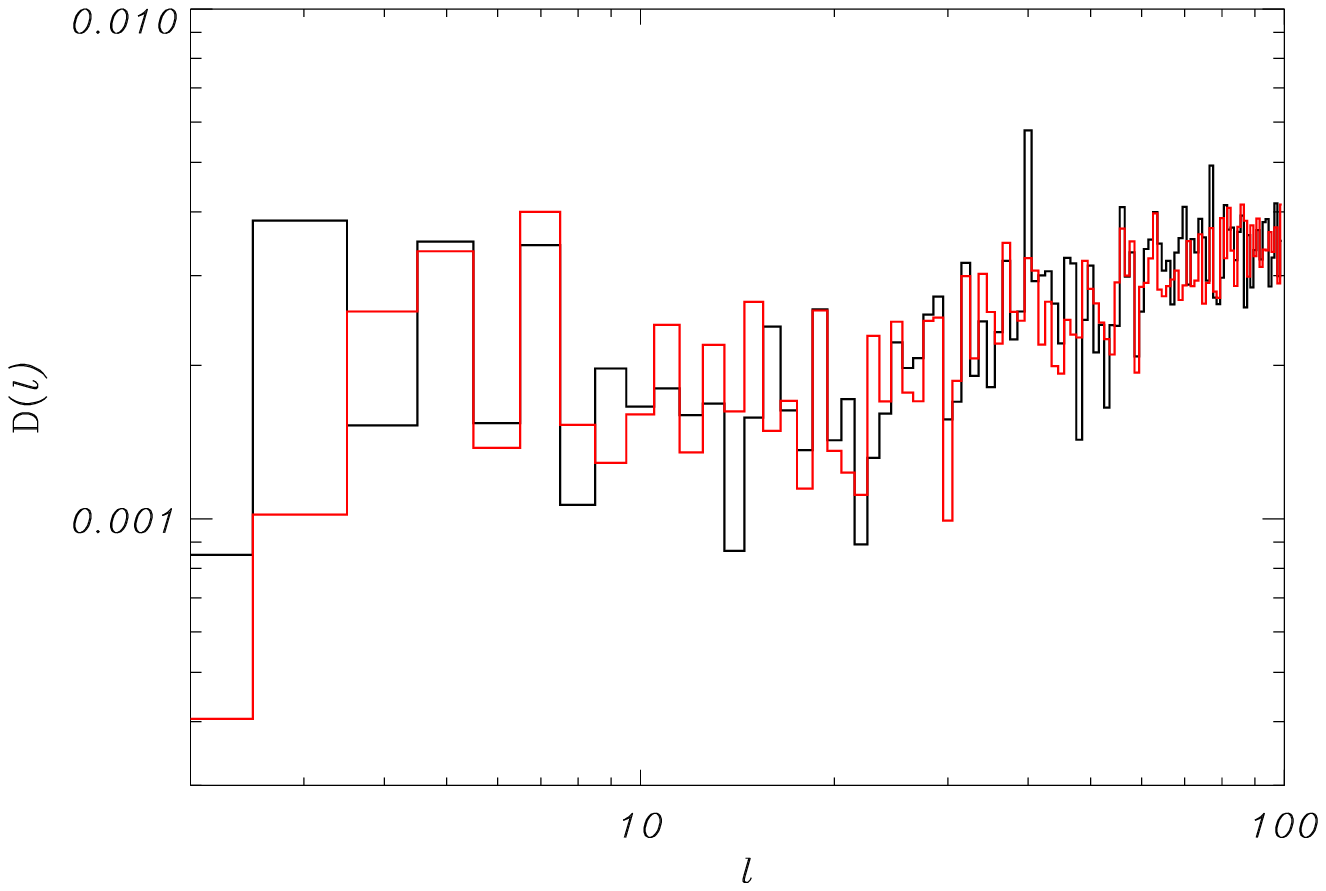}
\includegraphics[scale=0.6]{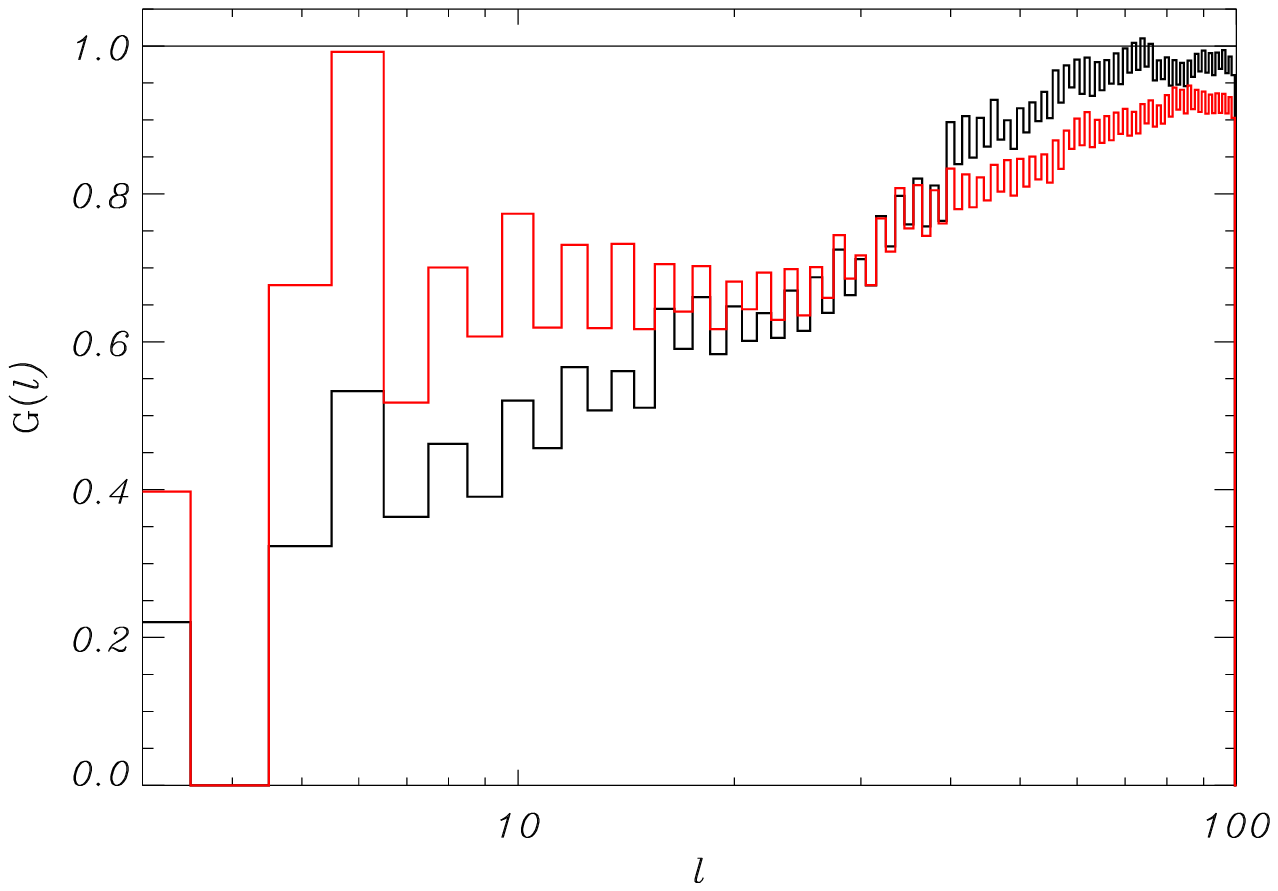}}
    \caption{The left panel: the power $D(l)$(in $\mu K^2$) for the galactic (black line) and ecliptic coordinates (red line). Right panel: the parity parameter $G(l)$ for the galactic and ecliptic coordinates (black and red respectively).}
    \label{ec}
  \end{center}
\end{figure}

The result of the investigation of Pearson's random walk in the domain of the mean angles of the phases, clearly indicate significant differences in the separation between the even and odd multipoles, in the Galactic and ecliptic system of coordinates. The separation of the CMB phases for even and odd multipoles, is very clear established in the Galactic coordinates, and it is less significant in ecliptic one. To confirm these predictions from the Pearson's random walk statistic, we perform a parity test $G(l)$ for the directional power $D(l)$, given by Eq(\ref{teg}). The results of the investigation are shown in Fig.\ref{ec} for the power $D(l)$ and corresponding parity parameter $G(l)$, for the case of galactic and ecliptic coordinates. As it follows from this figure, the dominance of odd modes over even ones occur in both system of coordinates, but the most significant departure from $G\sim 1$ take place at Galactic coordinates. The comparison of $G(l)$ test from 
Fig.\ref{ec} and $g(l)$-test from Fig.\ref{figure_par} show that for $2\le l\le 22$ the significance of these two tests is practically the same. 
It would be worth it to note, that unlike the parity test $g(l)$, defined on the power spectrum of the CMB signal, the directional G-test completely ignore the contribution of all $m=0$ modes, showing practically the same odd-even non-equality, as $g(l)$. 
This feature is important for the determination of possible sources of contamination of the primordial CMB signal (which we believe, is parity neutral ($g\sim 1$)) by exotic foreground residuals, such as the Kuiper belt. Recently the simplest model of the dust emissivity from the Kuiper belt (KBOE) was discussed in \cite{Burigana}, where emission of the KBOE has been modeled as a pure quadrupole component
with corresponding coefficients of the spherical harmonic decomposition $a^{KBOE}_{2,m=0}\neq 0$ and $a^{KBOE}_{2,m\neq 0}= 0$ in the ecliptic system of coordinates. In other words, the model only include contribution from the $m=0$-modes. According to our $G(l)$-test, this model for the foreground cannot contribute to the parity asymmetry, as the test exclude all $m=0$-modes, and we still find a parity asymmetry. Thus, to fit the observational data, this model needs significant modification, which would be investigated in a separate paper. Generally speaking, implementation of directional statistics in the space of phases and the power $D(l)$ seems to be very useful for determining the sources of parity asymmetry of the CMB.

\section{Discussion}
In this paper we have presented tests of the parity asymmetry of the Cosmic Microwave Background. We have tested the power spectrum of the CMB for differences in power of the even and odd multipoles, and found a power excess in the odd multipoles, for the entire range considered. \\
Additionally, we found peculiar alignments between mean angles for the odd and even multipoles. In galactic coordinates the separation between the $l$-values, for significant alignments, was 1, and for ecliptic the separation was 2 for most of the aligned multipoles.\\
Further, in order to investigate the parity asymmetry from the power spectrum, we looked at the corresponding mean angles, and took a closer look at the relationship between the even and odd multipoles in phase space via Pearson's Random Walk. We found, that there was a separation between the even and the odd multipoles, with strongest values for the galactic set of coordinates, where the statistical significance exceeded $3\sigma$. \\
We also checked the $R(l)$-values for various $l$, and found that the most significant contribution to the $R^-(l)$-values happened for $5\le l\le 30$. The even multipoles are on the other hand characterized by small values of $R^+(l)$.\\
Finally, we introduced the directional parity test, for various alignments of the coordinate system. We checked for galactic and ecliptic coordinates, and found, that the most significant departure from parity symmetry happened for the galactic system of coordinates, in accordance with the pearsons random walk test.

In conclusion, it is obvious, that there exists a significant parity asymmetry between the even and the odd multipoles, especially so for galactic coordinates. A nongaussianity in the signal from WMAP can have both instrumental and cosmological origin, and the coming data from the Planck Survey will help us greatly in determining if we see instrumental error in WMAP or have hints of new, undiscovered physics.

\section{Acknowledgments}
We thank the anonymous referee for constructive comments and remarks.
We acknowledge the use of the Legacy Archive for Microwave Background Data Analysis (LAMBDA). 
Our data analysis made the use of HEALPix \cite{Healpix_primer}.
This work is supported in part by Danmarks Grundforskningsfond, which allowed the establishment of the Danish Discovery Center.
This work is supported by FNU grant 272-06-0417, 272-07-0528 and 21-04-0355.

\end{document}